%% file: Paper_eng.tex
\documentclass[
 aps,
 11pt,
 final,
 notitlepage,
 oneside,
 twocolumn,
 nobibnotes,
 nofootinbib,
 superscriptaddress,
 centertags]
{revtex4}

\input{sao_cmd_author.tex}

\begin{document}
\selectlanguage{english}

\onecolumngrid
\begin{center}
\large \bfseries Spectroscopy of globular clusters in the dwarf spheroidal galaxy IKN
\end{center}
\begin{center}
\bfseries M.~E.~Sharina$^1$, V. V. Shimansky$^2$ \\
{\it $^1$ Special Astrophysical Observatory, Russian Academy of Sciences, \\
Nizhnii Arkhyz, 369167, Karachaevo-Cherkesskaya Republic, Russia \\
$^2$ Kazan (Volga Region) Federal University, ul. Kremlevskaya 18, Kazan, 420008 Russia} \\
\end{center}

\begin{center}
\begin{minipage}{\textwidth - 2cm}
\small 

Moderate resolution spectra of four globular clusters in the dwarf spheroidal galaxy IKN
obtained with the 6-m telescope of the Special Astrophysical Observatory have been used to determine the
radial velocities, ages, and metallicities of the clusters, and also to derive the first approximate estimates
of the abundances of Mg, Ca, and C. Cross-correlation with radial-velocity standards, fitting of the
observed spectra with model spectra, diagnostic diagrams based on the Lick absorption indices, and
comparison of the spectra and absorption indices with those of Galactic globular clusters are applied.
The integrated-light (IL) spectrum of the two bright clusters IKN4 and IKN5, which are close to the center of the
galaxy in projection on the celestial sphere, yields the heliocentric radial velocity
 $V_h= 38\pm30$~km/s, age $T=12.6\pm2$~Gyr, metallicity $[Fe/H]=-2.1\pm0.2$~dex and abundance of $\alpha$-process elements
 $[\alpha/Fe] \sim 0.5$~dex.
 The IL spectrum of the two weaker clusters IKN1 and IKN3, which are far from the center
of the galaxy, yields the radial velocity $V_h=-39 \pm 50$ km/s.
Despite of the low signal to noise ratio in the summary spectrum of IKN1 and IKN3, one can conclude from the comparison of 
the results of different used methods,  that IKN1 and IKN3 have seemingly the same age and metallicity as IKN4 and IKN5.
 According to the measured Lick indices 
H$_{\delta_{\rm F}}$ and H$_{\beta}$, the studied globular clusters in IKN have blue horizontal branches.
\end{minipage}
\end{center}
\begin{center}
\begin{minipage}{\textwidth - 2cm}
\small
{\bf Keywords}: {globular clusters: general---globular clusters: individual: 
[GPH2009]IKN-01, [GPH2009]IKN-03, [GPH2009]IKN-04, [GPH2009]IKN-05, NGC 6341, NGC~2419 --- dwarf galaxies: individual: IKN}
\end{minipage}
\end{center}

\twocolumngrid
\section{Introduction}
\label{intro}
The dwarf spheroidal galaxy IKN is a close neighbor of the spiral galaxy M~81, residing at a distance of
 3.75~Mpc from the Sun \cite{k06}. The main observational characteristics of IKN taken from \cite{k17} and
obtained in the present study are listed in Table~\ref{tab:ikn}: the distance to the galaxy, 
absolute magnitude in the Johnson-Cousins B filter $\rm M_B$  taking into account Galactic extinction, 
effective (or Holmberg) diameter $A_e$, average surface brightness in the band B inside the effective radius
 (columns 1--4, based on the data of \cite{k17}), the
maximum diameter of the galaxy based on the distribution of stars obtained with the Hyper Suprime-Cam of the 
Subaru telescope \cite{o15} and the heliocentric radial velocity derived in the present paper,
as the velocity of the brightest globular cluster in the galaxy (column 6). 
Looking ahead, we note that our measured velocity differs from the value obtained by Chiboucas et al. \cite{ch09} in 2009, $Vr=-140$~km/s .

Its large diameter and low surface brightness suggests IKN should be classified as an ultra-diffuse galaxy \cite{k17}.
It is difficult to estimate the true structural and photometric parameters of IKN due to a bright star 
projected onto its northern side. Even the position of the center of the galaxy is not known exactly \cite{t15}.

Seven candidate globular clusters were detected in IKN (\cite{g09},\cite{t15}), some of which are resolved
into stars in the Hubble space telescope (HST) images.
The names of the clusters in IKN identified in HST images by Georgiev et al. \cite{g09} are 
[GPH2009]~IKN-01, ..., [GPH2009]~IKN-07. We will refer for these as IKN1, ..., IKN~7 for brevity.
\begin{figure*}[]
 \setcaptionmargin{5mm} \onelinecaptionstrue \captionstyle{normal}
 \includegraphics[scale=0.56,angle=-90]{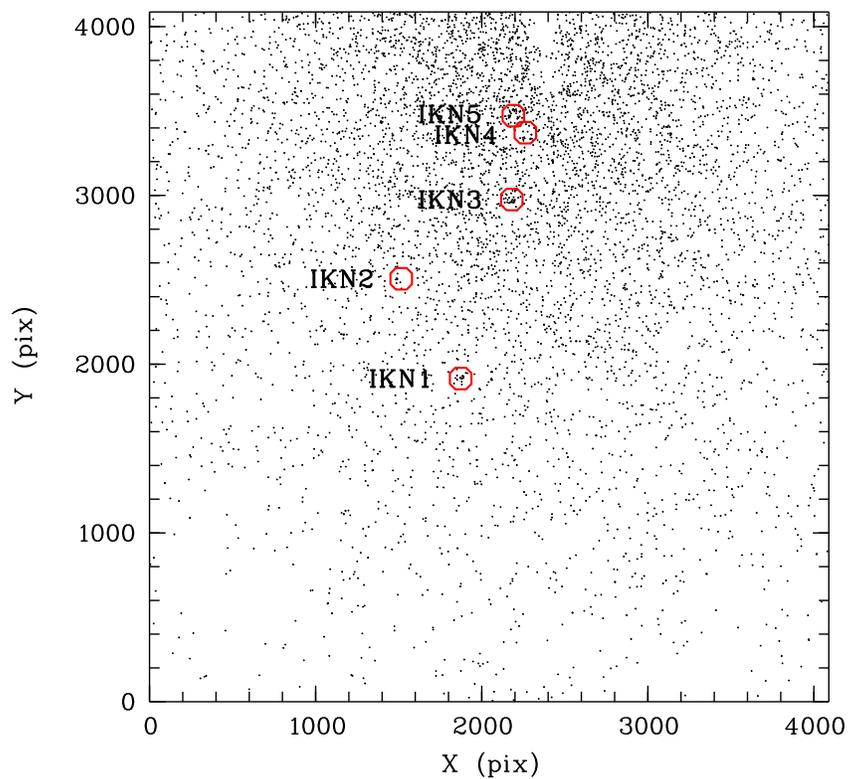}   
 \caption{Distribution of bright red giant stars in IKN in an image obtained from the HST stellar photometry \cite{k06}.
 The globular clusters are indicated. North is at the top and East to the left.}
\label{fig_dens}
\end{figure*}
\begin{figure*}[]
 \setcaptionmargin{5mm} \onelinecaptionstrue \captionstyle{normal}
 \includegraphics[scale=0.54]{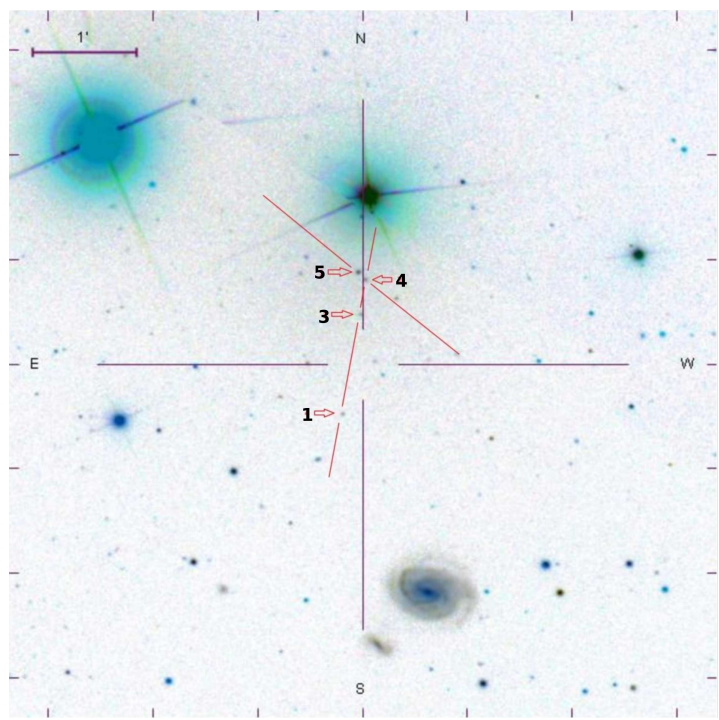}    
 \caption{Illustration of the positioning of the slits on the SDSS color image. The globular clusters are indicated.}
\label{slits}
\end{figure*}
\begin{table}[]
\setcaptionmargin{0mm} \onelinecaptionstrue \captionstyle{normal}
\caption{Observational characteristics of IKN.}
\label{tab:ikn}
\medskip
\begin{tabular}{|l|c|c|c|c|c|}
\hline   
$D,$ &   $M_B,$ & $\rm A_e,$& $SB_{Be},$      & $Diam$,& $V_h$,         \\  
Mpc  &  mag     &  kpc     & mag arcsec$^{-2}$ &   kpc &  km~c$^{-1}$   \\    \hline
3.75 &  -11.6 & 3.15      & 27.2           &  8.6  & 38$\pm 30$        \\  
\hline
\end{tabular}
\end{table}
Table~\ref{tab:propGCs} contains measurements of their structural and photometric parameters \cite{g09}
in the HST images. The columns of Table~\ref{tab:propGCs} contain: 
(1) the names of the objects following \cite{g09}; (2-3) the right ascension and declination
for epoch J2000.0; (4-5) the absolute magnitude and color; and (6) the half-light radius.
Georgiev et al. \cite{g09} provide the projected distances of the clusters from the center of the HST CCD
image. According to \cite{g09}, the brightest cluster, IKN5, is
more distant from the center of the galaxy than the remaining clusters.
According to stellar photometry results \cite{k06}, IKN5 resides in a region with
a greater stellar density than those near the other clusters (Fig.~\ref{fig_dens}).
Only the brightest red giants of the galaxy with Johnson-Cousins I magnitude brighter than I$=-5.5$~mag are shown 
in Fig.~\ref{fig_dens}. Note, however, that the detection results for stars in IKN at the northern edge 
of the frame could be affected by the presence of the bright star in that region.  
It's influence can be seen in Fig.~1 as an empty region in the stellar distribution.
 Given the spheroidal morphology of IKN and the observed distribution of the stellar density, 
it can be argued that IKN~5 is closer to the center of the galaxy than the other galactic clusters (see also \cite{t15}).

Based on an echelle spectrum with resolution R~8000 and a total exposure time of 2.3 hours taken with the Keck II telescope,
Larsen et al. \cite{l14} estimated  $\rm [Fe/H]=-2.1$~dex for IKN5,  as well as an age that is similar 
to those for most old Galactic globular clusters. They concluded that the number of stars in IKN5
is approximately equal to the number of metal-poor stars with $\rm [Fe/H]\sim-2.1$~dex in the galaxy itself.
\begin{table}[]
\setcaptionmargin{0mm} \onelinecaptionstrue \captionstyle{normal}
\caption{Characteristics of globular clusters in IKN from \cite{g09}.}
\label{tab:propGCs}
\medskip
\begin{tabular}{|l|c|c|c|c|c|c}
\hline
 Object  & $\alpha$(2000)    & $\delta$(2000)& $ M_V$   & $\!V\!-\!I$ & $r_h$ \\
         & hhmmss       &  grmmss   & mag    &  mag    & pc \\
\hline                                                                       
 IKN5    & 100805.5     & +682458   & $-8.47$  & 0.91      & 2.9   \\
 IKN4    & 100804.8     & +682454   & $-7.41$  & 0.94      & 2.0   \\
IKN3     & 100805.3     & +682434   & $-6.76$  & 1.09      & 14.8  \\
IKN1     & 100807.1     & +682337   & $-6.65$  & 0.91      & 6.6   \\
\hline
\end{tabular}
\end{table}

We performed spectroscopic observations of four globular clusters in IKN and determined their
ages, and the abundances of Fe, C, Mg and Ca. Section~\ref{data} describes the observational data
and the methods used to process them. In section~\ref{Spectroscopy} we analyze the spectra of globular clusters in IKN
by comparing them with data for Galactic globular clusters and stellar population models. A summary of our study
is presented in section~\ref{discussion}.

\section{Observational data and their reduction}
\label{data}
Globular clusters in IKN were observed at the 6-m telescope of the Special Astrophysical Observatory (SAO) using the SCORPIO-1 multi-mode 
focal reducer \cite{a05} in the spectroscopic mode with a long slit width 1". 
The observations were carried out in 2013-2015 on five dark nights with mostly clear sky or in the presence of weak cirrus
clouds, with various seeing. A log of the observations
is presented in Table~\ref{tab:log6m}. In addition to the observation date and the names of the clusters toward which the
spectrograph slit was pointed, the table lists the exposure time in seconds and the approximate Full Width
at Half Maximum (FWHM) of the stellar images in arcsec. The VPHG1200G grating was used for the
2013 observations, and the VPHG1200B grating for the 2014--2015 observations. A complete list of the
instrumental gratings is available at the SAO web-site\footnote{\url{https://www.sao.ru/hq/lon/SCORPIO/scorpio.html}}.
 The gratings we used have 1200 lines/mm, and provide a spectral resolution $FWHM\! \sim \!5.5 - 5$~\AA\ and 
 a dispersion 0.87 -- 0.88 \AA\ $\rm pix^{-1}$. 
The operative wavelength ranges are $\sim$3900-5500~\AA\ for VPHG1200B and
$\sim$4000-5700~\AA\ for VPHG1200G. The slit positions during the observations are shown in Fig.~\ref{slits}.
\begin{table}[]
\setcaptionmargin{0mm} \onelinecaptionstrue \captionstyle{normal}
\caption{Observational log.}
\label{tab:log6m}
\medskip
\begin{tabular}{|l|c|c|c|c|}
\hline   
Object      &  Data     & $T_exp$, c      &  $FWHM$," \\ \hline 
IKN5,4     & 14.03.2013& 2x600           &   3.5     \\ 
IKN5,4     & 14.03.2013& 8x900           &   2.5    \\ 
IKN3,1     & 15.03.2013&  2x900          &    1.3   \\ 
IKN3,1     & 15.03.2013&  4x1200         &    1.3   \\ 
IKN5,4     & 01.02.2014&  6x1200         &    2.6   \\ 
IKN3,1     & 01.02.2014&  6x1200         &    2.6   \\ 
IKN4,3,1   & 15.02.2015&  9x1200         &    1.5   \\ 
IKN5,4     & 18.02.2015&  6x1200         &    1.6   \\ 
IKN3,1     & 18.02.2015&  6x1200         &    1.6   \\ 
\hline
\end{tabular}
\end{table}

We used the spectra of the Galactic globular clusters NGC~6341 and NGC~2419 from our earlier studies (\cite{s13}, \cite{s18}),
 where the observations and reduction are described, as well as spectra from \cite{sch05}.
The observations of NGC~6341 and NGC~2419 were carried out with the 1.93-m
telescope of the Hault Provence observatory (OHP) using the CARELEC spectrograph
\footnote{\url{http://www.obs-hp.fr/guide/carelec/carelec-eng.shtml}}.

The reduction of the SAO 6-m telescope observations was similar to that described, e.g., in \cite{s18}. 
The long-slit spectra were reduced using the MIDAS (European Southern Observatory Munich Data Analysis System, \cite{b83}) 
and IRAF (Image Reduction and Analysis Facility, \cite{Tody}) software packages. The transformation of the observed spectra
into the wavelength scale by means of the identification of lines in spectra of a He-Ne-Ar lamp shows
that the mean uncertainty of the wavelength calibration is $\sim$0.16~\AA. 
One-dimensional spectra were extracted using the IRAF procedure {\it apsum}.
Sky emission lines were subtracted using the IRAF procedure {\it background}.
The difficulty of the spectra reduction process of these SAO 6-m telescope observations
was due to the necessity to sum spectra obtained in different observational datasets with different weather
and instrumental conditions to increase the signal-to-noise ratio ($\rm S/N$) 
and significantly improve the subsequent data analysis. 
During the combining the spectra, it was necessary to check whether the dependence of the resolution and the line spread function
(LSF) varied significantly from spectrum to spectrum. For this, we compared the wavelengths of
the sky emission lines with their catalog values and
constructed the LSF based on radial-velocity standards and spectra of the twilight sky 
and the observed globular clusters using the \textit{ULySS}\footnote{\url{ http://ulyss.univ-lyon1.fr}} 
 (University of Lyon Spectroscopic analysis Software, \cite{k08,k09}).

 \section{ANALYSIS OF THE SPECTRAL DATA}
\label{Spectroscopy}
\begin{figure*}[]
 \setcaptionmargin{5mm} \onelinecaptionstrue \captionstyle{normal}
\hspace{.5cm}
\begin{tabular}{p{0.9\textwidth}}
\hspace{2cm} (a) \\
 \includegraphics[bb= 72 240 540 523,clip,scale=0.9]{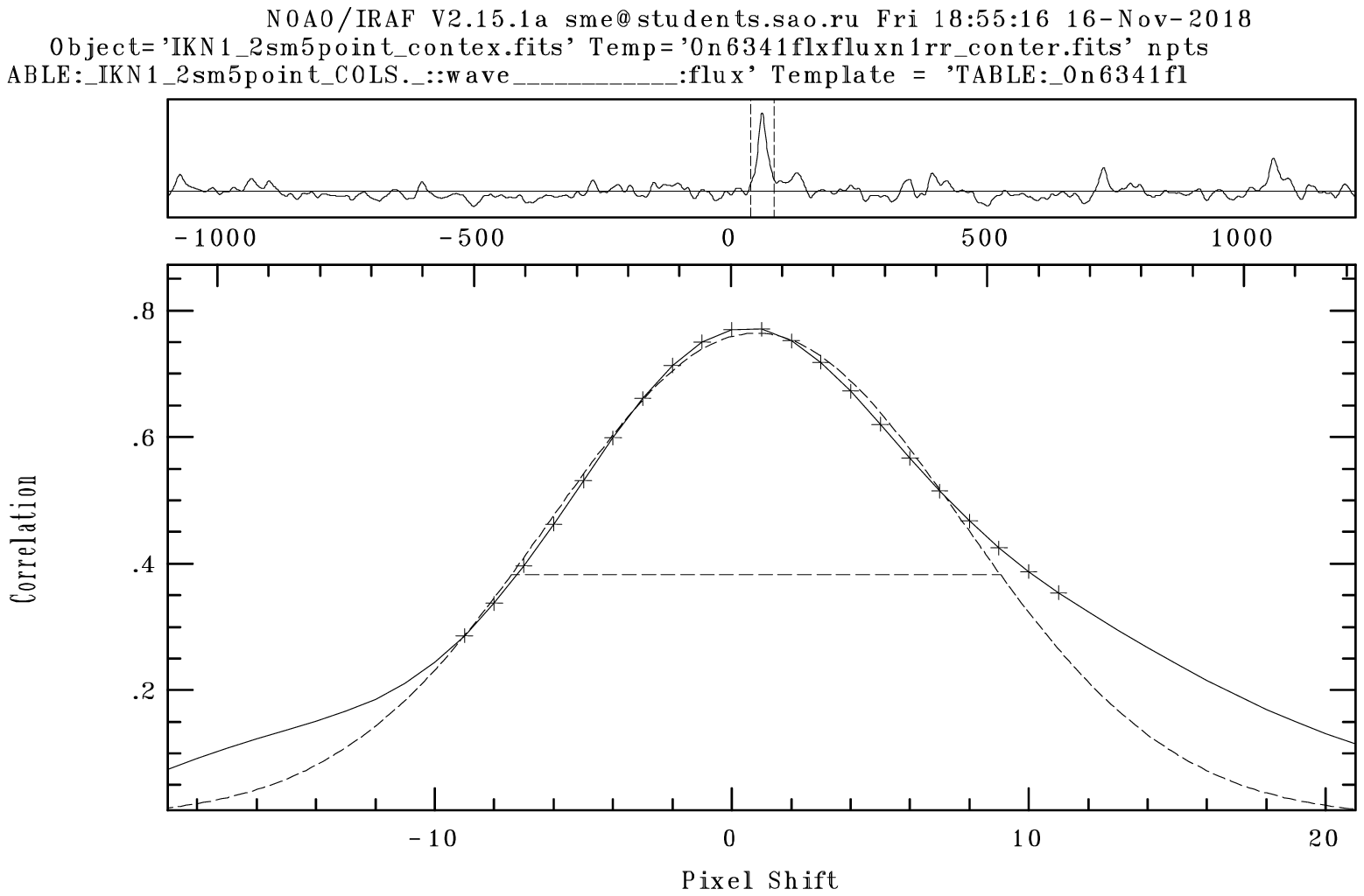} \\
\hspace{2cm} (b) \\
 \includegraphics[bb= 72 240 540 523,clip,scale=0.9]{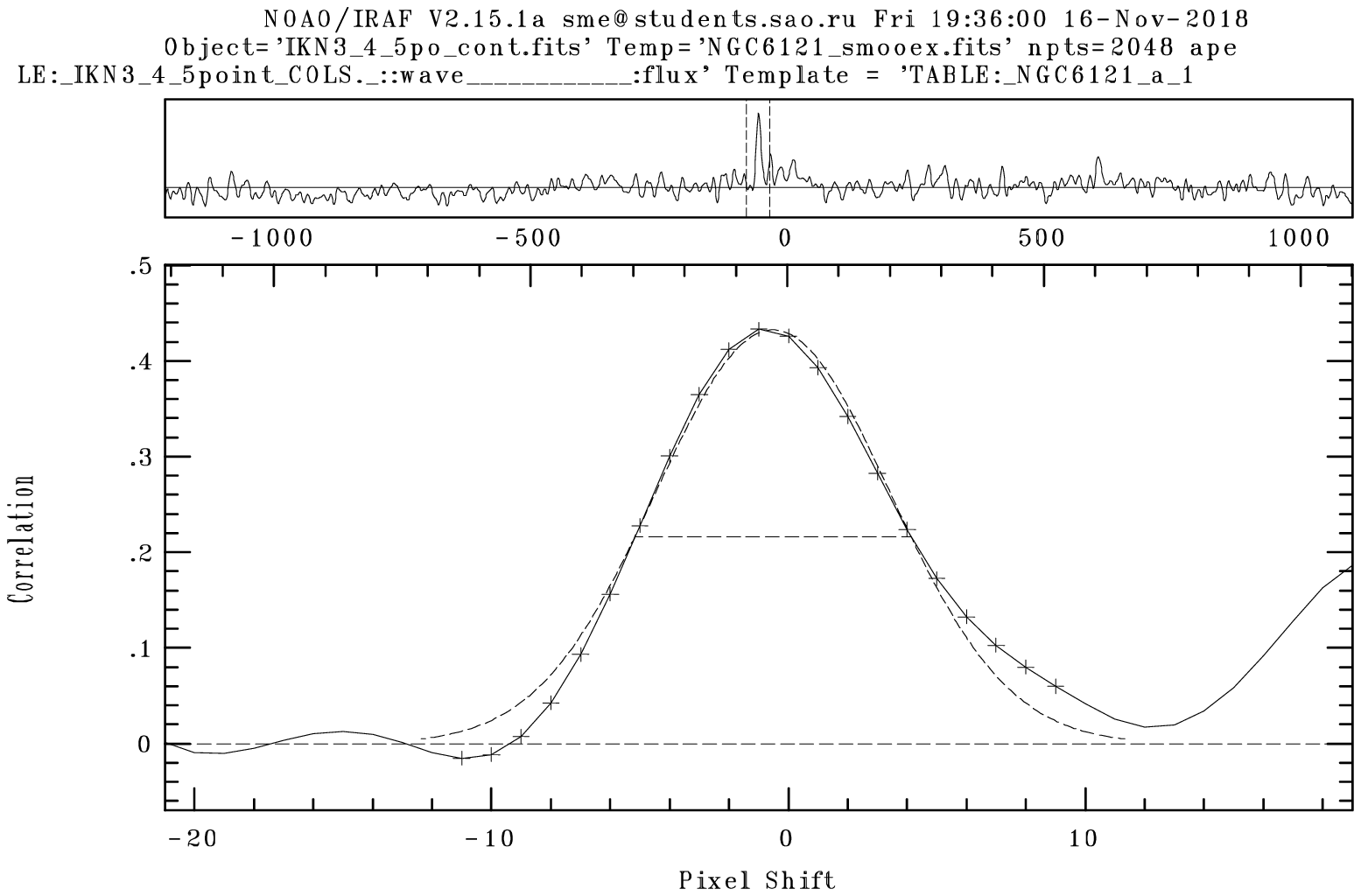} \\
\end{tabular}
\vspace{-0.2cm}
\caption{
Cross-correlation of the integrated-light (IL) spectra of IKN4+IKN5 (panel a) and IKN1+IKN3 (panel b) with spectra of the
twilight sky.  The top panels of each of the plots show the full cross-correlation functions. 
The vertical dashed lines mark the boundaries of the sections of the functions shown in the lower panels. 
The intensities of the cross-correlation peaks are plotted on the vertical axes. 
The offsets between the spectra of the object and of the radial-velocity standard in pixels (scales at the bottom of the figures) 
and in km/s ( scales at the top of the lower panels) are plotted along the horizontal axes in the lower panels. 
The dashed lines in the lower panels show the results of Gaussian fitting of the cross-correlation peaks.
}
\label{fig:fxcor}
\end{figure*}
\begin{figure*}[]
 \setcaptionmargin{5mm} \onelinecaptionstrue \captionstyle{normal}
\hspace{0.9cm}
 \begin{tabular}{p{0.9\textwidth}}
\hspace{1.5cm} (a) \\
 \includegraphics[scale=1.2,angle=0]{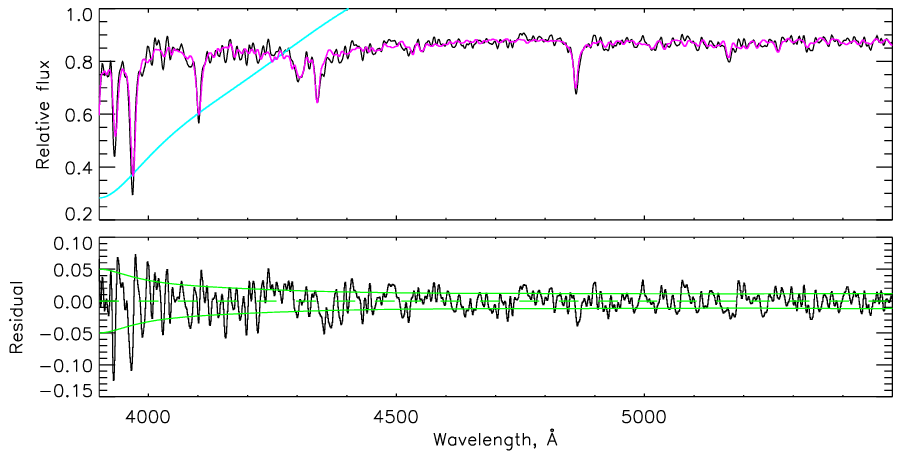} \\
\hspace{1.5cm} (b) \\
 \includegraphics[scale=1.2,angle=0]{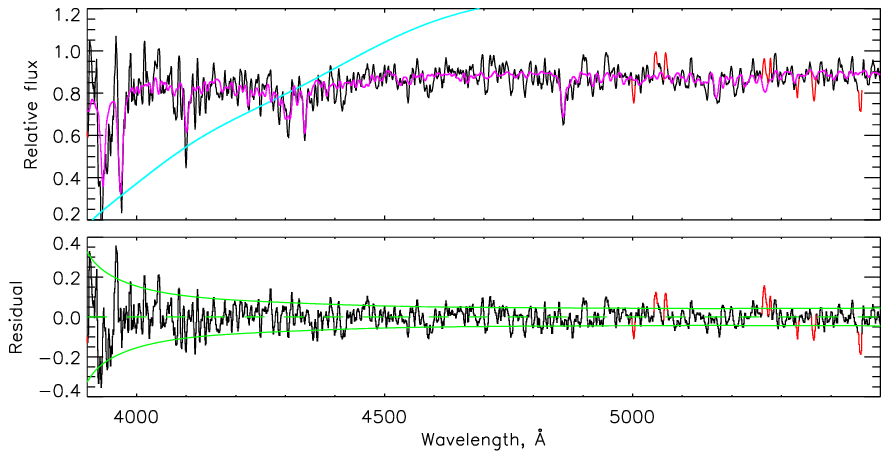} \\
\end{tabular}
\vspace{-0.2cm}
\caption{Integrated-light spectra  IKN4+IKN5 (Fig.~a) and IKN3+1 (Fig.~b)  with the model spectra for
NGC 6341 \cite{s18} (Fig.~a) and NGC~3201 \cite{sch05} (Fig.~b)  superposed (smooth curves). 
The results of dividing the spectra of objects by the corresponding model spectra are partially visible, shown by thin solid lines. 
The lower panels display the results obtained by subtracting the model spectra from the observed cluster spectra. 
The solid curves in the lower panels show the envelopes corresponding to 
to $\rm S/N \sim65$ for IKN4+IKN5 and $\rm S/N \sim18$ for IKN3+1. 
The dotted line shows the zero line. Parts of the cluster spectra significantly deviating from the model are highlighted.}
\label{fig:ulyssIKN4_5}
\end{figure*}

As a result of combining the spectra of the IKN globular clusters obtained on different nights, the
following one-dimensional spectra were obtained: the integrated-ight spectrum of the two globular clusters
IKN4 and IKN5, which are close to each other in projection onto the celestial sphere, and the spectra
of the two weaker clusters IKN1 and IKN3 (Fig.~\ref{slits}, Table~\ref{tab:propGCs}).
The $\rm S/N$  per pixel at the maximum sensitivity of the instrument at wavelength 5000\AA\ was $\rm S/N\sim65$ for IKN4+IKN5, 
and $\rm S/N\sim18$ for IKN1+IKN3. The spectra of IKN4 and IKN5, and of IKN1 and IKN3 were summed in order to increase the $\rm S/N$ 
through our analysis of the spectra of each of the objects obtained on different nights and to identify the stellar populations of
the clusters having similar velocities and properties.

The analysis of the spectra was performed as follows. We measured the radial velocities of the clusters
via a cross-correlation \cite{td79} with twilight-sky spectra using the {\it IRAF} procedure {\it fxcor}, 
and also fitting the obtained spectra with model spectra using the ULySS software {\it ULySS} \cite{k08,k09} and models from the PEGASE.HR  
 (Projet d'Etude des GAlaxies par Synthese Evolutive, \cite{lb04})  which employ the ELODIE library of stellar spectra \cite{ps01}.
Comparison of the observed spectra with the model spectra
and the spectra of Galactic globular clusters also enabled estimation of the ages and metallicities of the clusters.
Measurement of the Lick Absorption indices\footnote{\url{ http://astro.wsu.edu/worthey/html/index.table.html}}
\cite{Worthey94} and comparison with the analogous indices for Galactic globular clusters enabled an independent assessment 
of the ages and metallicities of the clusters, and also provides estimates of abundances of Mg, Ca, and C.

\subsection{Measurement of Radial Velocities}
\begin{table}[]
\setcaptionmargin{0mm} \onelinecaptionstrue \captionstyle{normal}
\caption{Radial velocities of the IKN clusters, derived from spectra obtained on different nights.}
\label{tab:radvel}
\medskip
\begin{tabular}{|l|c|c|}
\hline   
Object     &  Date     & $V_r$, km/s \\  \hline  
IKN4+IKN5     &14.03.2013 & 27$\pm$37   \\ 
IKN3+1     & 15.03.2013&  -20$\pm$54 \\ 
IKN4       & 01.02.2014&  72$\pm$43  \\ 
IKN3+1     & 01.02.2014&  -23$\pm$51 \\ 
IKN5       & 01.02.2014&  44$\pm$13  \\ 
IKN3+1     & 15.02.2015&  12$\pm$49  \\ 
IKN4       & 15.02.2015&  35$\pm$35  \\ 
IKN3       & 15.02.2015&   14$\pm$50 \\ 
IKN1       & 15.02.2015&  -38$\pm$52 \\ 
IKN5       & 18.02.2015&  14$\pm$46  \\ 
IKN4       & 18.02.2015&   2$\pm$50 \\ 
IKN3       & 18.02.2015& -21$\pm$55  \\ 
IKN1       & 18.02.2015& -13$\pm$50 \\ 
IKN1+IKN3     & 18.02.2015& -44$\pm$36  \\ 
\hline
\end{tabular}
\end{table}
The IL spectra of IKN1, IKN3, IKN4, and IKN5 were cross-correlated with spectra 
of the twilight sky on each observing night. The resulting measurements of the radial velocities of the clusters
according to the spectra obtained on different observational nights are listed in Table~\ref{tab:radvel}. 
Before the cross-correlation, we determined the continuum level in the spectra 
and divided the spectra by the continuum. We checked for possible systematical errors 
in the radial velocity estimates by measuring the wavelengths of 
sky emission lines in the spectra directly beneath the cluster spectra.
The resulting radial velocities of the objects obtained using this method are shown in Fig.~\ref{fig:fxcor}, 
 for the spectra averaged over all observing nights.
The measured radial velocities are: $+47$~km/s for IKN4+IKN5 and $-30$~km/s for IKN1+IKN3. 
Taking into account the heliocentric correction $-9$~km/s,
the heliocentric radial velocities are: $V_h=38$~km/s for IKN4+IKN5 and $V_h=-39$~km/s for IKN1+IKN3. 
We found no radial-velocity difference for IKN4 and IKN5 based on the spectra taken on different nights.
The velocity difference between IKN4+IKN5 and IKN1+IKN3 we have found may indicate that the
galaxy is rotating. However, confirmation of this conclusion will require additional spectral 
observations of the weak clusters.

The velocity uncertainty for IKN4+IKN5, $\sim$30~km/s, is less than the width of the cross-correlation peak in
Fig.~\ref{fig:fxcor}, since the velocity was measured independently over four observing nights.
The velocity uncertainty error for IKN1+IKN3 is $\sim$50~km/s, which corresponds to the width of the
cross-correlation peak in Fig.~\ref{fig:fxcor}. The velocity we measured for IKN5 is significantly 
different from the value derived by Chiboucas et al. \cite{ch09} from a medium-resolution spectrum 
with an exposure of 30 min
obtained using the FOCAS instrument mounted on the Subaru telescope: $Vr=-140$~km/s.

Similar results were obtained from a direct comparison of the observed cluster spectra with 
model spectra using the {\it ULySS} 
program \cite{k08,k09} with the PEGASE.HR models \cite{lb04}. The procedure  used for this  
comparison will be described in detail in the next section. 
A comparison of the spectra of IKN4+IKN5 and IKN1+IKN3 with the spectra of the Galactic globular clusters
NGC~6341 \cite{s18} and of NGC3201 \cite{sch05} is shown in Fig.~\ref{fig:ulyssIKN4_5}.

\subsection{Fitting of Models to the Observed Spectra}
\label{fullfit}
A direct comparison of the spectra of IKN4+IKN5 and IKN1+IKN3 with the spectra of Galactic globular clusters shows, 
that the intensities and profiles of the hydrogen and metal lines in the spectrum of IKN4+IKN5 best match the spectra of 
NGC~6341 \cite{s18} (Fig.~\ref{fig:ulyssIKN4_5}, upper panel). The spectrum of IKN1+IKN3 is most similar to the spectrum of 
NGC~3201 from \cite{sch05} (Fig.~\ref{fig:ulyssIKN4_5}, lower panel), for which
the metallicity and age are $[Fe/H] = -1.59 \pm 0.2$ and $T = 10.2 \pm 0.4$~Gyr \cite{r14}.
Comparisons of the spectra of other Galactic Globular clusters with the spectrum of IKN4+IKN5 
can be seen at the ftp site: \url{ftp://ftp.sao.ru/pub/sme/}. These show,
in particular, that the hydrogen and calcium Ca II H+K, 
G-band $\lambda \sim4300$\AA\ , and the region
of the molecule MgH 4980--5183\AA\ are best described by the spectrum of NGC~6341, suggesting that  
 the ages and chemical compositions of IKN4+IKN5 and NGC~6341 are similar.

We also compared the observed spectra with model spectra calculated using the
 PEGASE.HR \cite{lb04} code and the ELODIE \cite{ps01} library of stellar spectra. 
 The comparison with the model spectra was performed using the {\it ULySS} \cite{k08,k09} package.
A nonlinear minimization of the mean squared difference between the observed and model spectra was 
performed, varying the parameters of the latter in order to achieve the best fit to the observations.
The model parameters describe the properties of the stellar populations of the objects studied (age, metallicity,
stellar velocity dispersion projected onto the line of sight), as well as the influence of instrumental and other effects
distorting the observational data (absorption by dust, instrumental shifts, spectral-line broadening).
 The program minimized the $\chi^2$ residual between 
the observed data and model spectra convolved with the instrumental function of the spectrograph.
Multiplicative and additive components characterized by the Legendre polynomials were added to the model spectra. 
The spectrograph LSF was subject to a preliminary analysis for each observational night, using spectra
of standard stars and the twilight sky.

In agreement with the results obtained via a direct comparison of the spectra of the IKN
and Galactic clusters, the metallicity of IKN4+IKN5 determined from the comparison 
with the model spectra is $\rm [Fe/H]\sim-2.1\pm0.2$.
The metallicity of IKN1+IKN3 is higher: $\rm [Fe/H]\sim-1.6\pm0.45$.
The age of IKN4+IKN5 proved to be comparable to the age of the Universe $T=12\pm2$~Gyr. 
The age of IKN1+IKN3 is slightly younger $\sim7$~Gyr.
Since the spectrum of IKN1+IKN3 has a very low $\rm S/N$, its younger age and relatively 
high metallicity may result from 
the contribution of the night sky lines to the object's spectrum.
Due to low $\rm S/N$, a pixe-by-pixel comparison of the IKN1+IKN3 spectrum with the model and Galactic-cluster spectra 
should be treated with caution. As we will show in the following section, the Lick indices are preferable for analysis of the
IKN1+IKN3 spectrum.

\begin{table*}[]
\setcaptionmargin{0mm} \onelinecaptionstrue \captionstyle{normal}
\caption{Lick indices ($\lambda \leq5015$\AA\ ),
measured in the IL spectra of IKN4+IKN5, IKN1+IKN3, NGC~2419 and NGC~6341 (from top to bottom).}
\label{tab:lickind1}
\scriptsize
\begin{tabular}{lc|c|c|c|c|c|c|c|c|cc} \\ \hline                                                                                                                           
& H$_{\delta_{\rm A}}$ & H$_{\delta_{\rm F}}$ & CN$_1$         & CN$_2$         & Ca4227       &G4300        & Fe4531       & Fe4668         & H$_{\beta}$ & Fe5015       \\ 
&  (\AA)               & (\AA)                & (mag)          & (mag)          & (\AA)        & (\AA)        & (\AA)        & (\AA)          &  (\AA)       & (\AA) \\ \hline 
&3.77$\pm0.14$      &2.94$\pm0.14$           &-0.080$\pm0.001$&-0.054$\pm0.002$& 0.14$\pm0.06$& 1.74$\pm0.07$& 0.80$\pm0.09$ & 0.90$\pm0.10$ &2.64$\pm0.11$ & 0.87$\pm+0.12$ \\
&3.40$\pm0.90$      &2.84$\pm0.94$           &-0.104$\pm0.049$&-0.001$\pm0.012$& 0.36$\pm0.45$& 3.07$\pm0.48$& 1.57$\pm0.61$ &-0.05$\pm0.68$ &1.95$\pm0.70$ & 3.52$\pm+0.75$ \\
&3.11$\pm0.01$      &3.05$\pm0.01$         &-0.011$\pm0.0001$&-0.031$\pm0.0001$& 0.29$\pm0.003$& 1.98$\pm0.004$& 0.95$\pm0.01$ & 0.26$\pm0.01$ &2.19$\pm0.01$ & 0.70$\pm0.01$ \\
&3.78$\pm0.12$      &3.25$\pm0.13$           &-0.085$\pm0.001$&-0.058$\pm0.002$& 0.24$\pm0.06$& 2.40$\pm0.07$& 0.91$\pm0.09$ & 0.15$\pm0.10$ &2.20$\pm0.10$ & 1.54$\pm0.10$ \\
\hline
\end{tabular}
\end{table*}
\begin{table*}[]
\setcaptionmargin{0mm} \onelinecaptionstrue \captionstyle{normal}
\caption{Lick indices ($\lambda > 5015$\AA\ ),
measured in the IL spectra of globular clusters IKN4+IKN5, IKN1+IKN3, NGC~2419 and NGC~6341 (from top to bottom).}
\label{tab:lickind2}
\scriptsize
\begin{tabular}{lc|c|c|c|c|c|c|c|c|cc} \\ \hline                                                                                                                               
 & Mg$b$        & Mg$_1$          &  Mg$_2$        &  Fe5270       & Fe5335          & Fe5406         & Fe5782      & Na5895        & TiO$_1$         & TiO$_2$\\
&  (\AA)       & (\AA)            &  (mag)          &  (mag)       & (\AA)           & (\AA)          & (\AA)       & (\AA)         &  (mag)          & (mag)\\ \hline 
&0.85$\pm0.13$ &-0.005$\pm0.003$  & 0.033$\pm0.003$ &-0.16$\pm0.13$& 0.71$\pm0.13$   & 0.27$\pm0.13$  &   --        &   --          &    --           &     --         \\
&0.90$\pm0.81$ & 0.039$\pm0.021$  & 0.077$\pm0.021$ & 0.06$\pm0.82$& 0.44$\pm0.84$   & 1.26$\pm0.84$  &   --        &   --          &    --           &     --         \\
&0.47$\pm0.01$ & 0.002$\pm0.0002$ & 0.025$\pm0.0002$& 0.53$\pm0.01$& 0.61$\pm0.01$   & 0.33$\pm0.01$  &-0.01$\pm0.01$&1.04$\pm0.01$  &-0.006$\pm0.001$ &-0.002$\pm0.001$ \\
&0.12$\pm0.11$ & 0.015$\pm0.003$  & 0.024$\pm0.003$ & 0.44$\pm0.11$& 0.62$\pm0.11$   & 0.29$\pm0.11$  &0.11$\pm0.11$&1.39$\pm0.11$  & 0.034$\pm0.003$ & 0.005$\pm0.003$ \\
\hline
\end{tabular}
\end{table*}

\subsection{Lick index diagnostic diagrams}
\subsubsection{Method} 
\begin{figure*}[]
 \setcaptionmargin{5mm} \onelinecaptionstrue \captionstyle{normal}
\begin{tabular}{p{0.49\textwidth}p{0.49\textwidth}}
\hspace{-0.1cm}
\hspace{1.5cm} a & \hspace{2.5cm} b \\
\vspace{-1.cm}
 \includegraphics[angle=-90,scale=0.35]{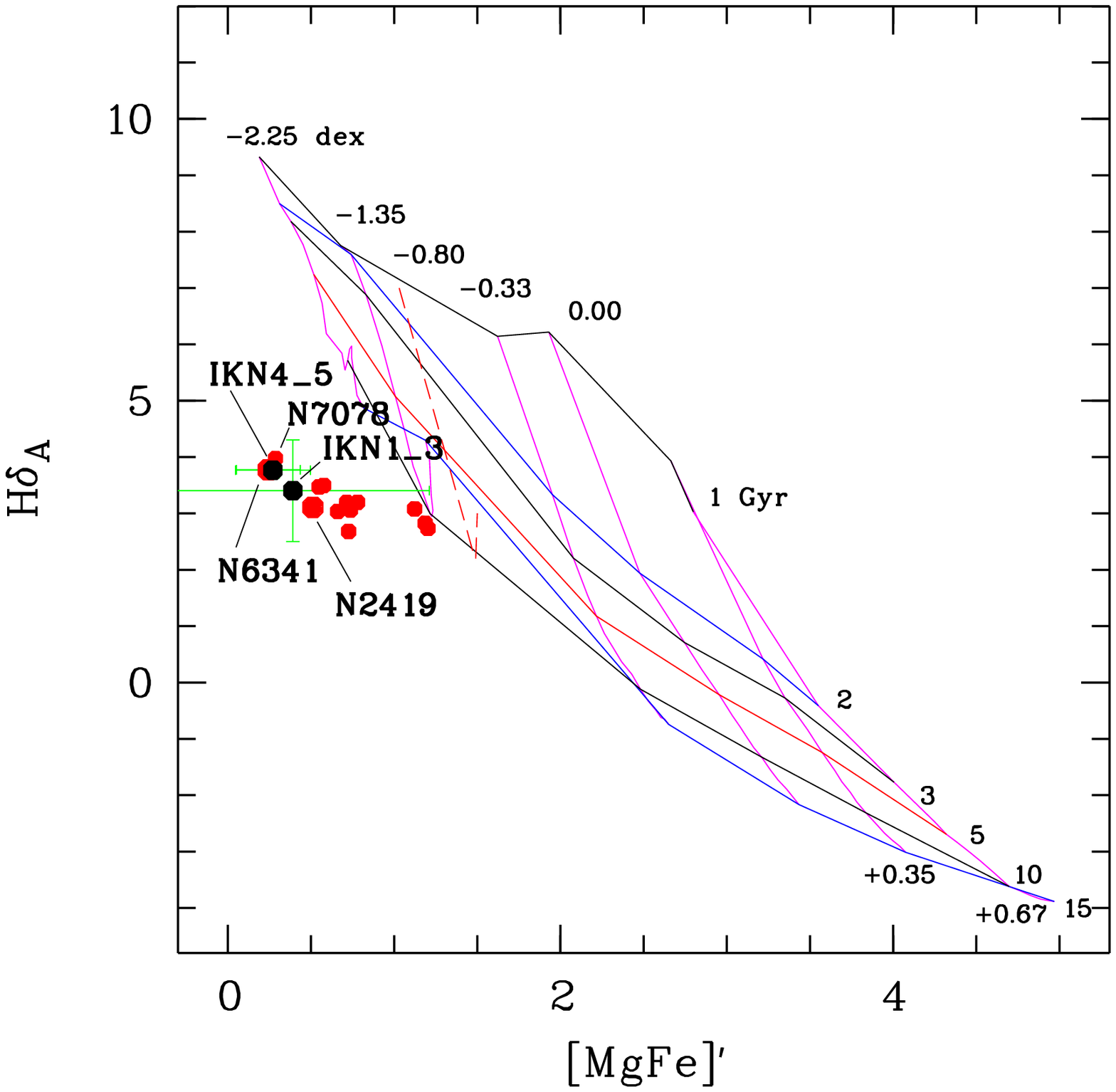} &
\vspace{-1.cm}
\hspace{0.4cm}
 \includegraphics[angle=-90,scale=0.35]{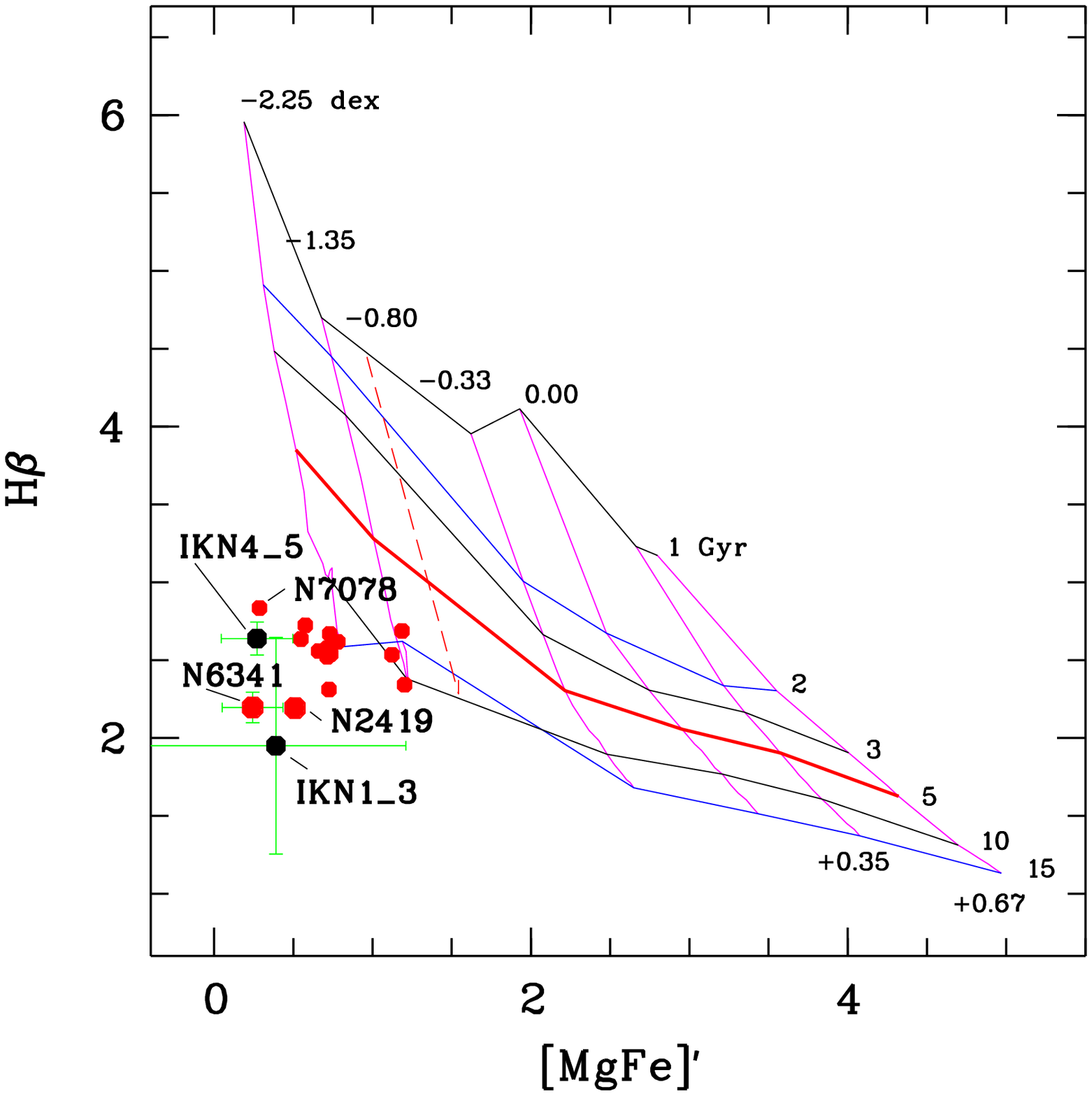} \\
\hspace{1.5cm} c & \hspace{2.5cm} d \\
\vspace{-1.cm}
 \includegraphics[angle=-90,scale=0.35]{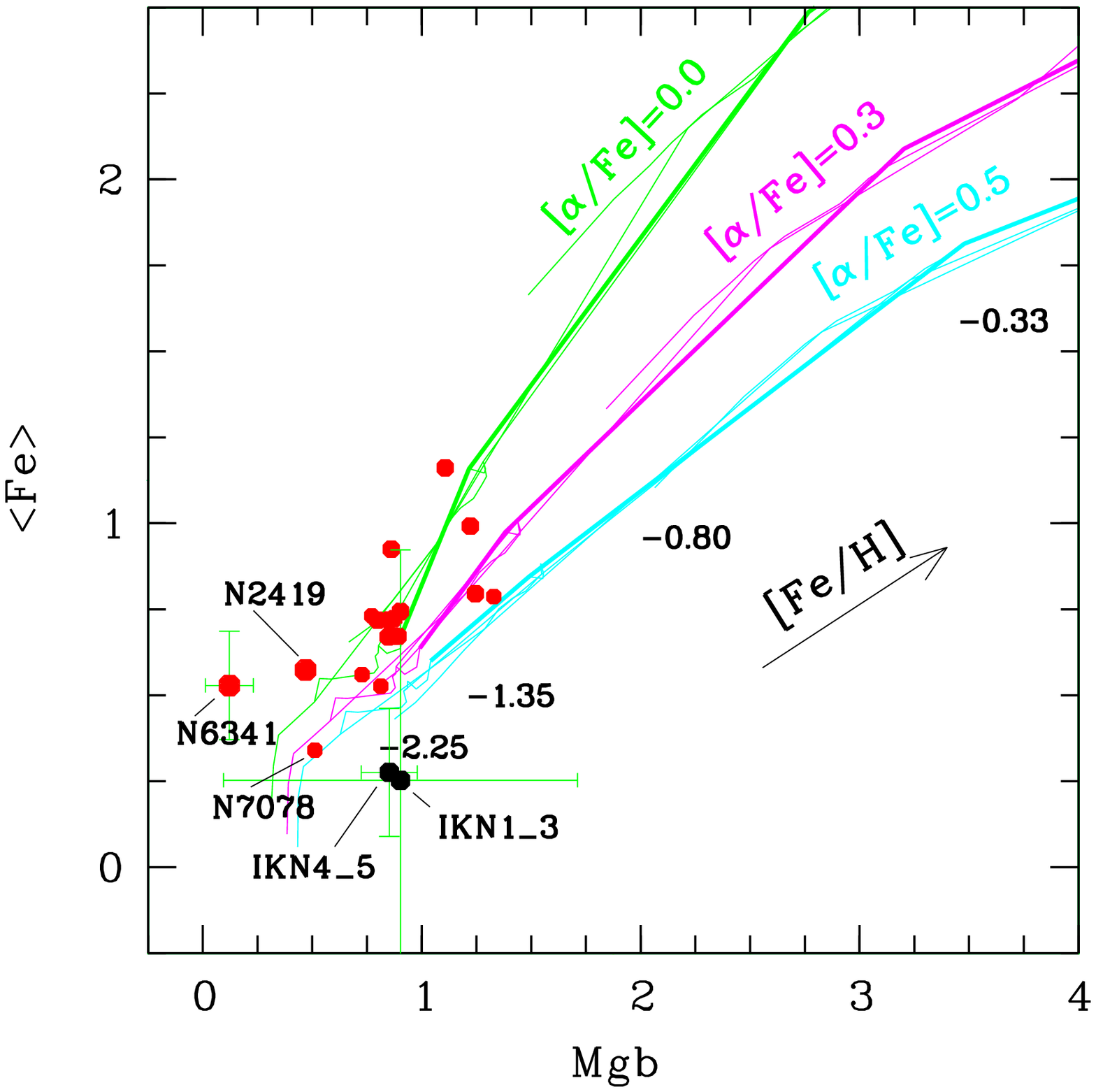} &
\vspace{-1.cm}
 \includegraphics[angle=-90,scale=0.35]{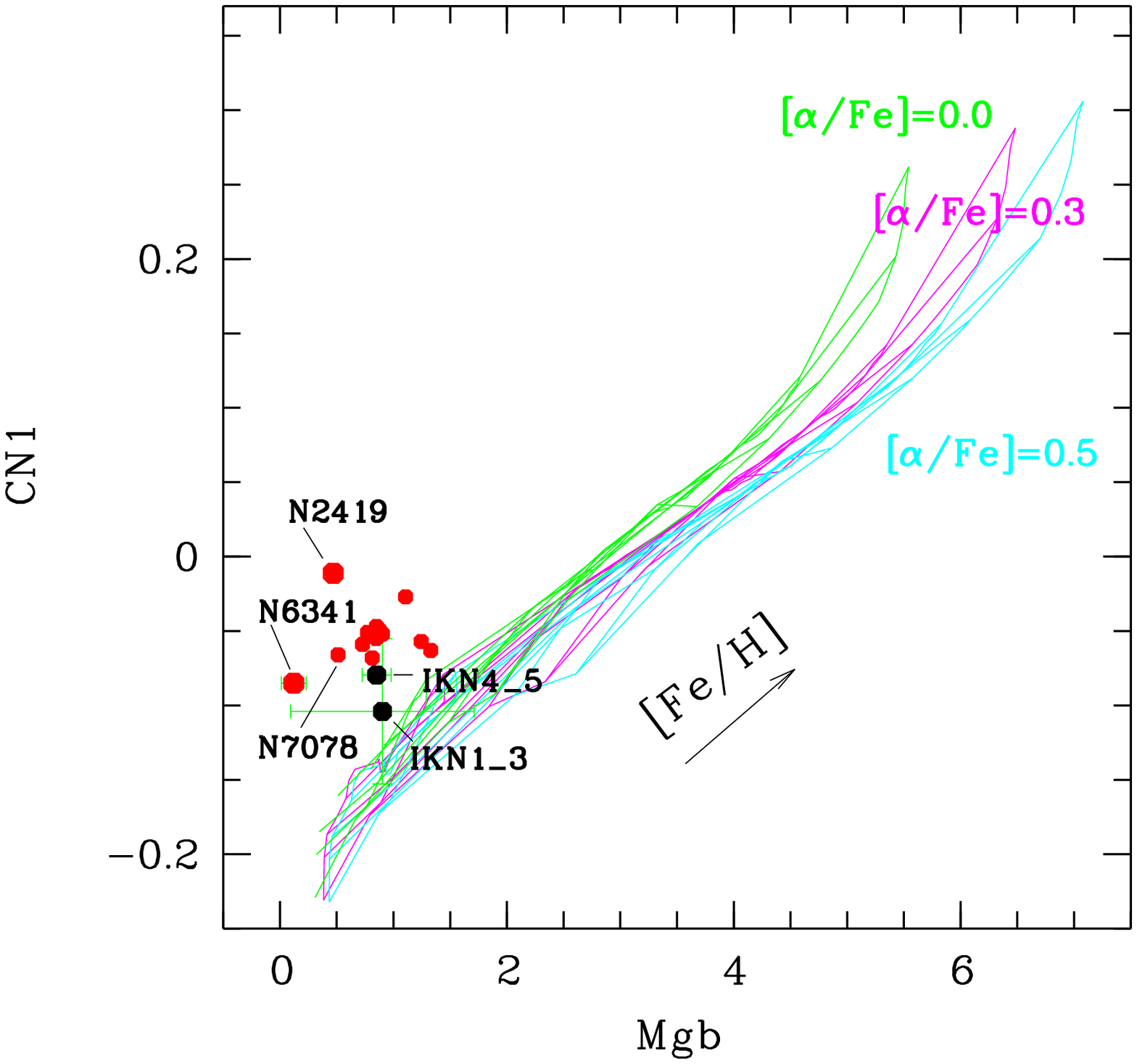} \\
\hspace{1.5cm} e & \hspace{2.5cm} f \\
\hspace{-0.3cm}
\vspace{-1.5cm}
 \includegraphics[angle=-90,scale=0.35]{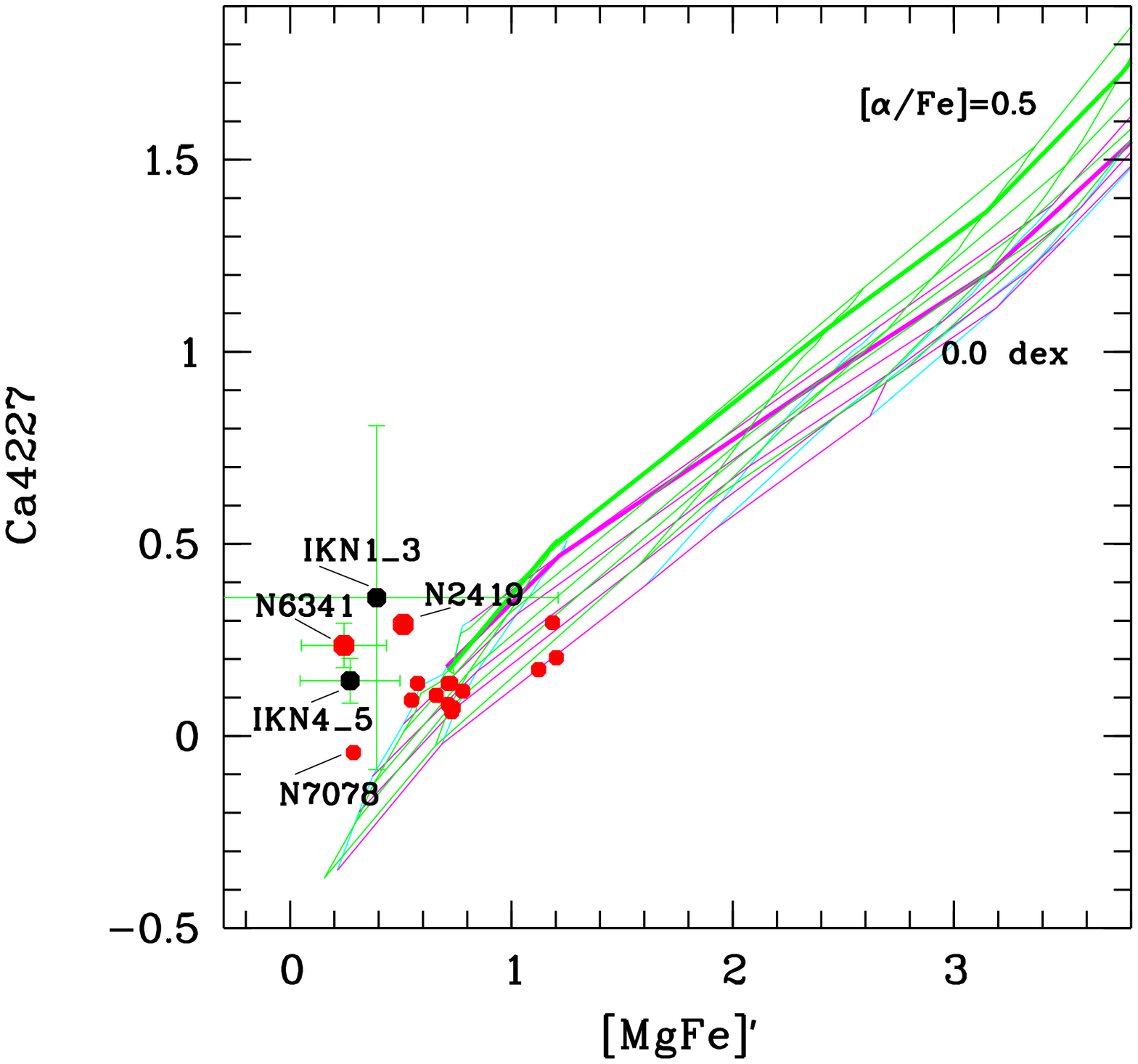} &
\hspace{0.4cm}
\vspace{-1.5cm}
 \includegraphics[angle=-90,scale=0.35]{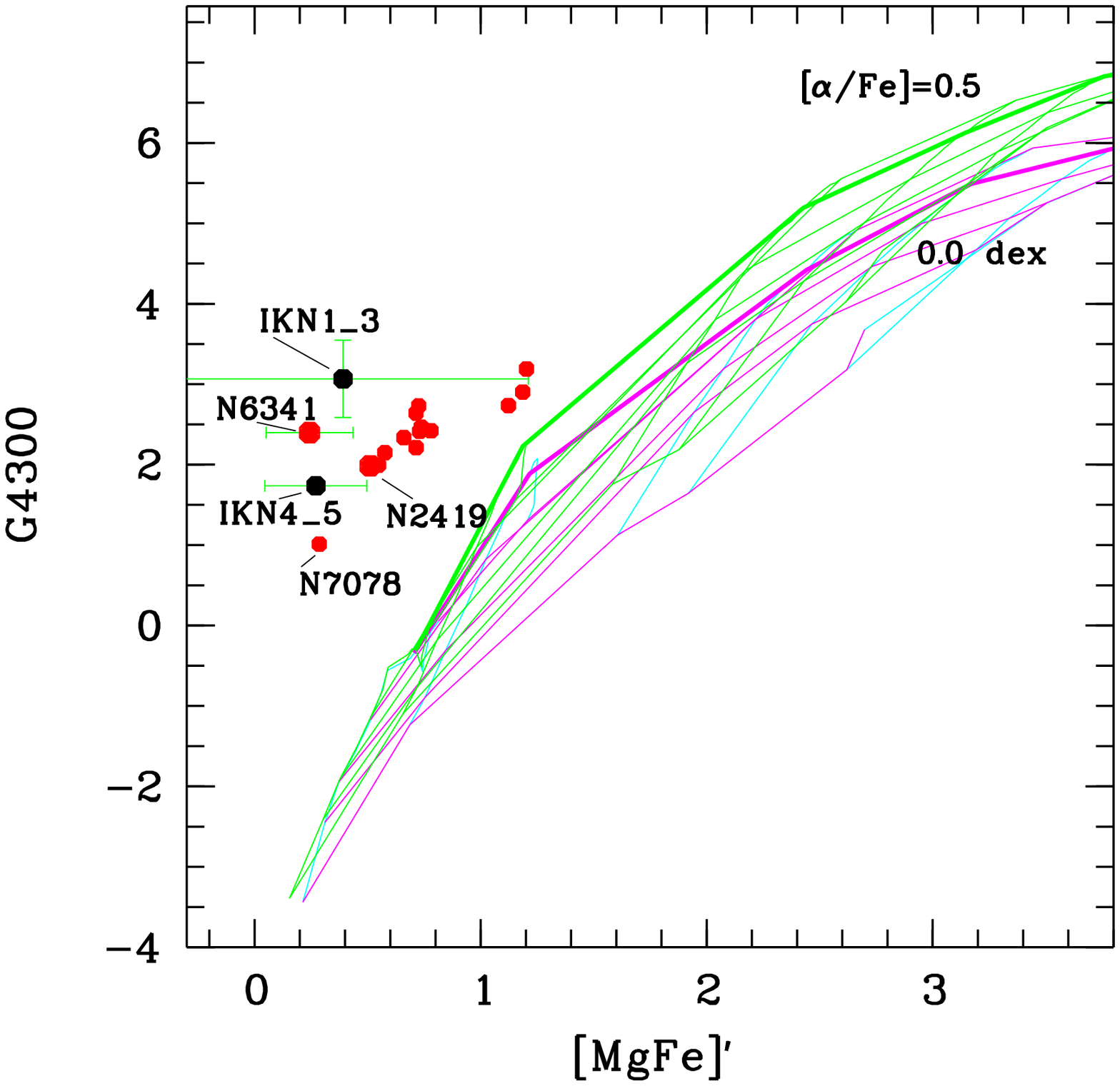} \\
\end{tabular}
\vspace{1.5cm}
 \caption{Lick index diagnostic diagrams. The Panels show (a), (b) age - metallicity diagrams,
(c) a metallicity - $\alpha$-process element abundance diagram, (d) a diagram of the indices of the molecular bands CN versus MgH, 
(e) and (f) the index centered at Ca~4227 \AA\ versus $\rm [Mg/Fe]'$ and the index CH (G4300) versus $\rm [Mg/Fe]'$. 
 Simple stellar population models (\cite{t03}, \cite{t04}) are shown by lines (see the text for details).}
\label{fig:lick}
\end{figure*} 
It is generally accepted in the literature, that the Lick system of
absorption-line indices\footnote{\url{ http://astro.wsu.edu/worthey/html/index.table.html}}
(\cite{b_84}, \cite{Worthey94}, \cite{w94}, \cite{wo97}, \cite{t98}) provides an effective tool for  determining
the ages and metallicities of globular clusters with ages $T\ge1$~Gyr.
The Lick indices are analogs of equivalent widths:
$ I(\lambda) = (1 - \frac{\int F_l(\lambda)\Delta \lambda)}{\int F_c(\lambda)\Delta \lambda)})\Delta \lambda$, where
$ \lambda$ is the wavelength, $ F_l$ the flux in the line and $ F_c$ the flux in the continuum in the vicinity of the line.
In addition to central wavelength regions centered on the strongest spectral features, 
the indices are characterized by
two areas of background-flux measurements to the right and to the left of the central area.
The measured Lick indies do not depend on the method used to correct the continuum in the spectra.
The typical widths of the central areas where the indices are measured is $20\div65$~\AA\ . Therefore, 
Lick indices are more efficient than a pixel-by-pixel approach to comparing 
the observed and model spectra, especially in the case of low $\rm S/N$.

We will now list the indices that are sensitive to the ages and abundances of some elements.
The values of the indices centered on the hydrogen lines mainly depend on the age:
H$_{\beta}$, H$_{\delta_{\rm A}}$, H$_{\delta_{\rm F}}$, H$_{\gamma_{\rm A}}$, H$_{\gamma_{\rm F}}$.
The metallicity is proportional to the indices centered on the iron lines:
 Fe5270,..., Fe4668, as well as
their combinations, e.g.,  $\rm\langle Fe\rangle=(Fe5270+Fe5335)/2$, 
[Mg/Fe]\arcmin$\rm=\sqrt{(Mgb\cdot(0.72\cdot Fe5270+ 0.28\cdot Fe5335))}$.
The value of the index Fe4668 is also defined by the carbon abundance, since the corresponding molecular lines 
 fall into the wavelength range of the index measurements. The index G4300 is most sensitive to the carbon abundance, 
while CN1 and CN2 are most sensitive to the carbon and nitrogen abundances together.
 The indices that are most sensitive to the abundance of $\alpha$-process elements are Mg$b$, Mg$_1$, Mg$_2$, and Ca4227. 

\subsubsection{Results}
Tables \ref{tab:lickind1} and \ref{tab:lickind2} present our measurements of the
Lick indices obtained using the procedure described in \cite{p04} 
for the IL spectra
of IKN4+IKN5 and IKN1+IKN3, as well as the spectra of NGC~2419 and NGC~6341  \cite{s13, s18}.
Fig.~\ref{fig:lick} shows diagnostic diagrams for the Lick indices. The lines connect the model indices from
(\cite{t03}, \cite{t04}) for equal age, or metallicity. These models
for simple stellar populations (\cite{t03}, \cite{t04}) contain 
Lick indices calculated for various ages, metallicities and $\rm [\alpha/Fe]$ values. The black filled circles
Fig.~\ref{fig:lick} show the
data for IKN4+IKN5 and IKN1+IKN3 and the red filled circles the data for NGC~2419 and NGC~6341  
(the objects are labeled in the figure), and also other Galactic globular clusters 
with metallicities $\rm [Fe/H]<1.3$ from \cite{Schiavon12}.
The objects from \cite{Schiavon12} are not labeled in the plot panels, however,  
it is clear that they are separated into two groups.
One group of three clusters contain NGC~6218, 6235 and 3201. The other group contains nine objects with 
$\rm [Fe/H] \leq -1.6$: NGC~1904, NGC~2298, NGC~5286, NGC~5946, NGC~5986, NGC~6254, NGC~6333, NGC~6752, and NGC~7078.
Only one cluster from \cite{Schiavon12} (NGC~7078) has a metallicity $\rm [Fe/H]< -2.0$~dex; this cluster is also 
labeled in Fig.~\ref{fig:lick}. 

The $\rm S/N$ in the spectrum of IKN4+IKN5 is much higher than the $\rm S/N$ in the spectrum of IKN1+IKN3, and
the uncertainties in the measured indices for IKN4+IKN5 are accordingly much lower.
It is clear from tables \ref{tab:lickind1} and \ref{tab:lickind2} and Fig.~\ref{fig:lick}, 
that despite their relatively high errors, all the indices for IKN1+IKN3 have values close to those for IKN4+IKN5, and this 
cannot be a random result.

The ratio H$_{\delta_{\rm F}}/$H$_{\beta}=1.11$ for IKN4+IKN5, while this ratio for IKN1+IKN3, NGC~2419 and NGC~6341 
varies in the range 1.4-1.5. This indicates that 
all these clusters have blue horizontal branches, according to \cite{Sch04}. 
It was shown in \cite{Sch04} that H$_{\delta_{\rm F}}/$H$_{\beta} \ge1.1$ for low-metallicity
Galactic globular clusters with blue horizontal branches.

Using the method described in \cite{s09} (their Section~3.1.1. and Appendix~B) and the models \cite{t03, t04},
we found the ages of IKN4+IKN5 and IKN1+IKN3 to be $\sim$12.6~Gyr. The uncertainty in the age of IKN4+IKN5 is $\sim20\% $,
while the uncertainty in the age of IKN1+IKN3 reaches $\sim$50\% due to the low $\rm S/N$ in this spectrum.
The metallicities of IKN4+IKN5 and IKN1+IKN3 appear to be approximately equal, $\rm [Fe/H]\sim-2.0$. 
 However, the dispersion of this value 
is much higher for IKN1+IKN3 ($\sigma \sim 0.2$), than for IKN4+IKN5 ($\sigma \sim 0.04$).

The diagnostic age-metallicity diagrams in Figs.~\ref{fig:lick}a and \ref{fig:lick}b show
that the hydrogen-line indices for IKN4+IKN5 and IKN1+IKN3 are close to those of the three mentioned Galactic clusters with
metallicities $\rm [Fe/H]\leq-2.0$. Table~\ref{tab_evolGGCs} summarizes ours and literature estimates of the metallicities and ages  
for NGC~6341, NGC~7078, NGC~2419 and five clusters from \cite{s17} 
with the mean metallicity, $\rm[Fe/H]\!\sim\!-1.6$~dex (NGC~1904, NGC~5286, NGC~6254, NGC~6752, and NGC~7089). 
The average age for the clusters with $\rm [Fe/H]\leq-2.0$ is 12.7~Gyr. Our age estimates for
NGC~6341 and NGC~7078 and metallicity estimate for NGC~6341 \cite{s18} based on their IL spectra 
differ from the literature values obtained from color-magnitude diagrams (Table~\ref{tab_evolGGCs}).
This may due to contamination of the spectra by the emission of background stars.
The average metallicity of NGC~6341 and NGC~7078 derived from an analysis of color-magnitude diagrams \cite{VandenBerg} is $\rm[Fe/H]\!\sim\!-2.34$,
which is close to our estimates for IKN4+IKN5, obtained using Lick indices and fitting of the observed spectra using model spectra.
\begin{table*}
\setcaptionmargin{0mm} \onelinecaptionstrue \captionstyle{normal}
\caption{
Age (T) in Gyr, metallicity, abundances of C, Mg and Ca, 
and $\alpha$-process elements, as the mean of the Ca and Mg abundances (in dex), 
for NGC~6341, NGC~7078, and NGC~2419 having medium-resolution spectra similar to the spectrum of IKN4+IKN5.}
\label{tab_evolGGCs}
\begin{tabular}{|l|c|c|c|r|c|c|}    \hline \hline
NGC      & T         & [Fe/H]  &  [C/Fe]   & [Mg/Fe]  & [Ca/Fe]   & [$\rm \alpha/Fe$] \\  \hline\hline
6341     & 12.75$^4$ &-2.35$^4$& -0.41$^5$   & 0.11$^5$   & 0.16$^5$    &   0.14$^5$     \\
         & 11.20$^1$ &-1.80$^1$& -0.30$^1$   &-0.25$^1$   & 0.00$^1$    &  -0.13$^1$     \\   \hline                                                              
7078     & 12.75$^4$ &-2.33$^4$& -0.30$^6$   & 0.36$^6$   & 0.31$^6$    &  0.34$^6$     \\
         & 14.10$^1$ &-2.35$^1$& -0.15$^1$   &-0.23$^1$   & 0.15$^1$    &  -0.04$^1$     \\  \hline                                                            
2419     &    --     &-2.12$^7$& 0.15$^7$    & 0.30$^7$   & 0.15$^7$    &  0.23$^7$      \\
         & 12.60$^2$ &-2.25$^2$& 0.10$^2$    &-0.05$^2$   & 0.10$^2$    &  0.03$^2$      \\   \hline                                 
5 GCs    & 12.60$^3$ &-1.73$^3$& -0.03$^3$   & 0.27$^3$   & 0.18$^3$    &  0.23$^3$      \\
         & 12.75$^3$ &-1.60$^3$&  0.61$^{6(\!n\!=\!3)}$& 0.40$^3$& 0.29$^3$ &  0.35$^3$     \\         
\hline         
\end{tabular}
\scriptsize
\begin{tabular} {rp{90mm}}
For five globular clusters with $\rm[Fe/H]\!\sim\!-1.6$ from \cite{s17} (last two rows) the parameter values averaged over the group are given.\\[-5pt]
Last row contains literature data for red giants in these clusters (see \cite{s17} and references therein). The data were taken from: \\[-5pt]
$^1$ Sharina et al. (2018) \cite{s18}; $^2$ Sharina et al. (2013) \cite{s13}; $^3$ Sharina et al. (2017) \cite{s17}; $^4$ VandenBerg et al. (2013) \cite{VandenBerg};\\[-5pt]
$^5$ Meylan et al. (2001)\cite{Meylan}; $^6$ Roediger et al. (2014)\cite{r14}; $^7$ Cohen et al. (2011) \cite{c11}. \\[-5pt]
\end{tabular}
\end{table*}

The abundances of $\alpha$-process elements in IKN4+IKN5 and IKN1+IKN3 derived as described in \cite{s09},
by comparing the Lick indices of the clusters with the models from \cite{t03, t04} are $\rm [\alpha/Fe]\sim0.5$~dex.
This estimate is only approximate, since the model sequences for different values of $\rm [\alpha/Fe]$ are close for low metallicities. 
The metallicity versus $\alpha$-process element abundance plot in Fig.~\ref{fig:lick}c shows that the data points for IKN4+IKN5 and IKN1+IKN3
are positioned to the right and below the model with $\rm [\alpha/Fe]=0.5$, in contrast to the other clusters with
$\rm [Fe/H]\sim-2.3$~dex, for which the data points are positioned to the left and above the model with $\rm [\alpha/Fe]=0.0$~dex.
This indicates that IKN4+IKN5 and IKN1+IKN3 have a high abundance of $\alpha$-process elements compared to other clusters with
$\rm [Fe/H]\sim-2.3$, which have $\rm [\alpha/Fe]<0.0$~dex.

Table~\ref{tab_evolGGCs} shows that the $[Mg/Fe]$ values for objects with $\rm [Fe/H]\leq-2.0$ derived in our studies 
based on the IL medium-resolution spectra of Galactic clusters 
(\cite{s18}, \cite{s13}, \cite{s17}) are systematically lower,
than the literature values derived for cluster red giants using high-resolution 
spectrographs (\cite{VandenBerg}, \cite{Meylan}, \cite{r14}, \cite{c11}).
This circumstance was discussed in \cite{s17} in the context of the presence of multiple stellar populations in the clusters.
When we compare abundances from the literature with the data derived from the Lick-index diagnostic diagrams,
we must  first and foremost rely on the results of studying the IL spectra of clusters. 

The Ca4227 index for IKN4+IKN5 (Fig.~\ref{fig:lick}e) is 
slightly higher than the corresponding indices for NGC~6341 and NGC~2419.
According to our data \cite{s18} for NGC~6341 and NGC~2419, the mean $\rm [Ca/Fe]$ value in Table~\ref{tab_evolGGCs} is $0.05$~dex. 
Thus, a comparison with the data for the elemental abundances for low-metalicity 
Galactic clusters with similar spectra indicates 
that the lower limit for the abundance of $\alpha$-process elements in IKN4+IKN5 and IKN1+IKN3 is
$\rm [\alpha/Fe] =([Mg/Fe]+[Ca/Fe])/2 \sim 0.2$.
This is in reasonable agreement with our rough estimate of $\rm [\alpha/Fe]$ in IKN4+IKN5 and IKN1+IKN3 
based on the Lick indices, ($\rm [\alpha/Fe]\sim0.5$).

The values of the CN1 index for IKN4+IKN5 and IKN1+IKN3 are close to those for 
NGC~6341 and NGC~7078 (Fig.~\ref{fig:lick}d),
and are only slightly below the mean value of this index for the group of five Galactic clusters with $\rm [Fe/H]\sim -1.6$~dex.
The G4300 index (CH band $\lambda \sim 4300$~\AA\ ; Fig.~\ref{fig:lick}f) for IKN4+IKN5 
is close to those of NGC~7078, NGC~6341, and NGC~2419.
In general, the average G4300 indices for 
low-metallicity ($ \rm[Fe/H]<-2.0$) and higher-metallicity clusters 
in the corresponding diagram in Fig.~\ref{fig:lick} are similar.
In Table~\ref{tab_evolGGCs}, the average $[C/Fe]$ value derived from the IL spectra of the Galactic clusters 
(\cite{s13}, \cite{s18}, \cite{s17}) is approximately $\sim-0.1$ for all the objects. 
Taking into account the similarity of the spectra for IKN4+IKN5 and NGC~6341 in the region of the G band (Section~\ref{fullfit}), 
and the known  $[C/Fe]$ value for NGC~6341 (Table~\ref{tab_evolGGCs}), this suggests that 
$[C/Fe]\sim-0.3$~dex for IKN4+IKN5.
Thus, according to our rough estimate, the $[C/Fe]$ value for IKN4+IKN5 ranges from $\sim0.1$ to $-0.3$.

\section{Conclusions}
\label{discussion}

Using observations obtained with the SCORPIO-1 multimode focal reducer \cite{a05} in the prime focus 
of the SAO 6-m telescope, we have acquired IL spectra for several globular clusters in the dwarf spheroidal galaxy IKN: 
IKN4+IKN5 (total $S/N\sim65$ per pixel at 5000\AA), and IKN1+IKN3 (total $S/N\sim18$ per pixel at 5000\AA).
The radial velocities of the clusters proved to be appreciably different from the value, -140 km/s, found for IKN5 in \cite{ch09}
based on a medium-resolution spectrum with an exposure of 30 min.
We found that IKN4 and IKN5 have very similar heliocentric radial velocities, on average $V_h= 38\pm30$~km/s,
while the corresponding value for the IL spectrum of the clusters IKN1 and IKN3 is $V_h=-39 \pm 50$~km/s. 
The difference in the velocities of these pairs of clusters in IKN with similar properties of their stellar populations may correspond 
to the value of the velocity dispersion of stars in the galaxy, or can serve as a signal for the galaxy's rotation. 
Additional observations are necessary to confirm the detection of the rotation of IKN and to determine the velocity dispersion of stars in IKN.

We have estimated the metallicity, age and $\rm [\alpha/Fe]$ value for IKN4+IKN5, by
comparing the Lick absorption indices measured for these clusters with the model indices \cite{t03,t04} and
the indices for Galactic globular clusters, and also by directly comparing the cluster spectra with the model spectra \cite{lb04}
and the spectra of Galactic globular clusters: $[Fe/H]=-2.1\pm0.2$, $T=12.6\pm2$~Gyr and $[\alpha/Fe]=0.2\div0.5$~dex. 
The [Fe/H] values we have obtained are in agreement with the metallicity of IKN5 derived by Larsen et al. \cite{l14}, based on
a high-resolution echelle spectrum of IKN5.
Considering the similarity of the Lick indices measured for the IL spectrum of IKN4+IKN5 and those of
Galactic clusters with similar metallicities, we have made a first rough estimate of the abundances of Mg, Ca and C 
in the spectrum of IKN4+IKN5: $0.3<[Mg/Fe]<0.5$~dex, $[Ca/Fe]\sim 0.05$~dex and $0.1<[C/Fe]<0.3$~dex.

Based on the similarity of the majority of the corresponding Lick indices for IKN4+5 and IKN1+IKN3, measured in the 
IL spectra, we have concluded that these objects have similar metallicities and ages.
A pixel-by-pixel comparison of the spectrum of IKN1+IKN3 with models computed using 
PEGASE.HR \cite{lb04} and the ELODIE \cite{ps01} library of stellar spectra
in the {\it ULySS} package \cite{k08,k09} gives a higher metallicity and younger age for IKN1+IKN3:
$\rm [Fe/H]\sim-1.6\pm0.45$~dex and $T\sim 7$~Gyr. The uncertainty in the age of IKN1+IKN3 obtained
using this method is 50\%.
A direct comparison of the observed spectrum with the spectra of Galactic globular clusters indicates that the spectrum of 
IKN1+IKN3 is similar to the spectrum of NGC3201 which has metallicity $\rm [Fe/H]=-1.59\pm0.2$ and age $T=10.2\pm0.4$ \cite{r14}.
Thus we find that the age and metallicity of IKN1+IKN3 are confined to the ranges
$\rm7\le T \le12.6$~Gyr, $\rm 1.6\le[Fe/H]\le2.1$~dex.

We have identified signatures of the presence of blue horizontal branch stars in IKN5, IKN4, IKN3, and IKN1 
using the results by Schiavon et al. \cite{Sch04} and
the ratio between the Lick index values $H_{\delta_{\rm F}}$ and $H_{\beta}$ in the IL spectra of IKN4+IKN5 and IKN1+IKN3:
$H_{\delta_{\rm F}}/H_{\beta}>1.1$. \\

{\Large Acknowledgements}

The study was supported by the grant RFBR 18-02-00167.
V.V.Sh. acknowledges the funding from the Kazan State University  
 (a subsidy for completion of state contract in the sphere of
science 3.9780.2017/8.9).

\end{document}

%% file: sao_cmd_author.tex
\def\squareforqed{\hbox{\rlap{$\sqcap$}$\sqcup$}}

\def\sq{\ifmmode\squareforqed\else{\unskip\nobreak\hfil
\penalty50\hskip1em\null\nobreak\hfil\squareforqed
\parfillskip=0pt\finalhyphendemerits=0\endgraf}\fi}

\def\arcmin{\hbox{$^\prime$}}

\def\utw{\smash{\rlap{\lower5pt\hbox{$\sim$}}}}

\def\udtw{\smash{\rlap{\lower6pt\hbox{$\approx$}}}}

\def\diameter{{\ifmmode\mathchoice
{\ooalign{\hfil\hbox{$\displaystyle/$}\hfil\crcr
{\hbox{$\displaystyle\mathchar"20D$}}}}
{\ooalign{\hfil\hbox{$\textstyle/$}\hfil\crcr
{\hbox{$\textstyle\mathchar"20D$}}}}
{\ooalign{\hfil\hbox{$\scriptstyle/$}\hfil\crcr
{\hbox{$\scriptstyle\mathchar"20D$}}}}
{\ooalign{\hfil\hbox{$\scriptscriptstyle/$}\hfil\crcr
{\hbox{$\scriptscriptstyle\mathchar"20D$}}}}
\else{\ooalign{\hfil/\hfil\crcr\mathhexbox20D}}%
\fi}}


































%% file: Paper_eng.bbl
\begin{thebibliography}{199}
\bibitem{k06}
Karachentsev I.D., Dolphin A., Tully R.B., Sharina M., et al., Astron. J. {\bf 131}, 1361 (2006).
\bibitem{k17}
Karachentsev I. D., Makarova L. N., Sharina M. E., Karachentseva V. E., Astrophys. Bull. {\bf 72}, 376 (2017).
\bibitem{o15}
S. Okamoto, N. Arimoto, A. M. N. Ferguson, et al.,  Astrophys. J. Lett. 809, L1 (2015).
\bibitem{ch09}
K. Chiboucas, I.D. Karachentsev, R. B. Tully, Astron. J. {\bf 131}, 3009 (2009).
\bibitem{t15}
A. Tudorica, I. Y. Georgiev, and A. L. Chies-Santos, Astron. and Astrophys. {\bf 581}, 84 (2015).
\bibitem{g09}
I. Y. Georgiev, T. H. Puzia, M. Hilker, and P. Goudfrooij, Monthly Notices Royal Astron. Soc. {\bf 392}, 879 (2009).
\bibitem{l14}
S. S. Larsen, J.P. Brodie, D.A. Forbes, J. Strader, Astron. and Astrophys. {\bf 565}, 98 (2014).
\bibitem{a05}
V. L. Afanasiev and A. V. Moiseev, Astronomy Letters 31, 194 (2005).
\bibitem{s13}
M. E. Sharina, V. V. Shimansky, and E. Davoust, Astronomy Reports {\bf 57}, 410 (2013).
\bibitem{s18}
M. E. Sharina, V. V. Shimansky, and D. A. Khamidullina, Astrophys. Bull. {\bf 73}, 337 (2018).
\bibitem{sch05}
R. P. Schiavon, J. A. Rose, S. Courteau, and L. A. MacArthur, Astrophys. J. Suppl. 160, 163 (2005).
\bibitem{b83}
 K. Banse, P. Crane, P. Grosbol, et al., The Messenger {\bf 31}, 26 (1983).
 \bibitem{Tody}
D. Tody, in Astronomical Data Analysis Software and Systems II, 
ed. R. J Hanisch, R. J. V. Brissenden, and J. Barnes (San Francisco, CA: ASP), ASP Conf. Ser. {\bf  52}, 173 (1993).
\bibitem{k08}
M. Koleva, P. Prugniel, P. Ocvirk, D. Le Borgne, C. Soubiran, Monthly Notices Royal Astron. Soc. {\bf 385}, 1998 (2008).
\bibitem{k09}
M. Koleva, P. Prugniel, A. Bouchard, Y. Wu, Astron. and Astrophys. {\bf 501}, 1269 (2009).
\bibitem{td79}
Tonry, J., Davis, M. Astron. J.{\bf 84}, 1511 (1979).
\bibitem{lb04}
Le Borgne D., Rocca-Volmerange B., Prugniel P., Lancon A., Fioc M., Soubiran C., Astron. and Astrophys. {\bf 425}, 881 (2004).
\bibitem{ps01}
Prugniel P., Soubiran C., Astron. and Astrophys. {\bf 369}, 1048 (2001).
\bibitem{Worthey94}
G. Worthey, S. M. Faber, J.J. Gonzalez, and D. Burstein, Astrophys. J. Suppl. Ser. {\bf 94}, 687 (1994).
\bibitem{r14}
J.C. Roediger, S. Courteau, G. Graves, and R.P. Schiavon, Astrophys. J. Suppl. {\bf 210}, 10 (2014).
\bibitem{b_84}
D. Burstein, S. M. Faber, C. M. Gaskell, N. Krumm, Astrophys. J.  {\bf 287}, 586 (1984).
\bibitem{w94}
G. Worthey, Astrophys. J. Suppl. Ser. {\bf 95}, 107 (1994).
\bibitem{wo97}
G. Worthey \& D. L. Ottaviani, Astrophys. J. Suppl. Ser. {\bf 111}, 377 (1997).
\bibitem{t98}
S. C. Trager, G. Worthey, S. M. Faber, D. Burstein, J. J. Gonzalez, Astrophys. J. Suppl. Ser. {\bf 116}, 1 (1997).
\bibitem{p04}
T. H. Puzia, et al., Astron. and Astrophys. {\bf 415}, 123 (2004).
\bibitem{t03}
D. Thomas, C. Maraston, R. Bender, Monthly Notices Royal Astron. Soc. {\bf 343}, 279 (2003).
\bibitem{t04}
D. Thomas, C. Maraston, A. Korn, Monthly Notices Royal Astron. Soc. {\bf 351}, L19 (2004).
\bibitem{Schiavon12}
R. P. Schiavon, N. M. Caldwell, H. P. Heather, S. Courteau, L.A. MacArthur, G. J. Graves, Astron. J. {\bf 143}, 14 (2012).
\bibitem{Sch04}
R. P. Schiavon, J.A.~Rose, S. Courteau, L.A. MacArthur, Astrophys. J. {\bf 608}, L33 (2004).
\bibitem{s09}
M. Sharina, E. Davoust, Astron. and Astrophys. {\bf 497}, 65 (2009).
\bibitem{s17}
M. E. Sharina, V. V. Shimansky, and A. Y. Kniazev, Monthly Notices Royal Astron. Soc. {\bf 471}, 1955 (2017).
\bibitem{VandenBerg}
Don A. VandenBerg, K. Brogaard, R. Leaman, and L. Casagrande, Astrophys. J. {\bf 775}, 134 (2013).
\bibitem{Meylan}
G. Meylan, A. Sarajedini, P. Jablonka, S. G. Djorgovski, T. Bridges, R. M. Rich, , Astron. J {\bf 122}, 830 (2001).
\bibitem{c11}
J.G. Cohen, W. Huang, E.N. Kirby, 2011, ApJ 740, 60
\end{thebibliography}
